\documentclass[12pt]{article}
 \usepackage{amsfonts,latexsym,amssymb,amsthm}
 \usepackage{graphicx}

   \newcommand{\field}[1]{\mathbb{#1}}
   \newcommand{\rz}{\field{R}}
   \newcommand{\cz}{\field{C}}

   \newcommand{\dzd}[1]{\frac{\partial^2}{\partial #1^2}}

   \newcommand{\openset}{\mathcal{O}}
   \newcommand{\Hilbert}{\mathcal{H}}
   \newcommand{\feld}{\mathcal{F}}

 \newtheorem{theorem}{Theorem}[section]
 \newtheorem{definition}[theorem]{Definition}
 \newtheorem{exa}[theorem]{Example}
 \newtheorem{lem}[theorem]{Lemma}
 \newtheorem{cor}[theorem]{Corollary}
 \newtheorem{pro}[theorem]{Proposition}
 \newtheorem{rem}[theorem]{Remark}

\begin{document}

\title{On the local structure of the Klein-Gordon field on curved spacetimes}

\author{Alexander Strohmaier}

\date{\small Universit\"at Leipzig,
Institut f\"ur theoretische Physik,
Augustusplatz 10/11, D-04109 Leipzig, Germany\\
E-mail:  alexander.strohmaier@itp.uni-leipzig.de\\
WWW: http://www.physik.uni-leipzig.de/\~{}strohmai/}

\maketitle
\noindent
\begin{abstract}
  \noindent
  This paper investigates wave-equations on spacetimes with a metric which is locally
  analytic in the time. We use recent results in the theory of the non-characteristic Cauchy
  problem to show that a solution to a wave-equation vanishing in an open set
  vanishes in the ``envelope'' of this set, which may be considerably larger
  and in the case of timelike tubes may even coincide with the spacetime itself.
  We apply this result to the real scalar field on a globally hyperbolic
  spacetime and show that the field algebra of an open set and its envelope
  coincide.
  As an example there holds an analog of Borchers'
  timelike tube theorem for such scalar fields and hence, algebras
  associated with world lines can be explicitly given.
  Our result applies to cosmologically relevant spacetimes.
\end{abstract}
\maketitle

\section{Introduction}

 Since the canonical quantization procedure of fields requires a
 "frequency splitting" with respect to the time, the absence of a
 timelike symmetry group in general curved spacetimes causes
 severe problems (see \cite{Wald,Fulling:1989nb}) in the construction of
 quantum fields. The nature of these problems becomes particularly
 clear in the algebraic formulation of quantum field theory
 (\cite{Haag:1992hx}), where
 an algebra of observables $\mathcal{A}(\openset)$ is associated to each
 region $\openset$ in spacetime. 
 Specializing to the
 Klein-Gordon field the local $C^*$-algebras $\mathcal{A}(\openset)$ can be constructed in a
 similar manner as in Minkowski spacetime (see \cite{Dimock:1980hf})
 and the problem reduces to finding the class of physical states.
 This is usually achieved by choosing a physically preferred state, the vacuum.
 The construction then gives a net of local von Neumann algebras
 the normal states being the physical ones. We favour the $C^*$-algebraic approach
 since our result is most easily stated using the local von Neumann algebras.
 It is also common to use $*$-algebras  or the Borchers-Uhlmann-algebra of
 test functions (\cite{Borchers:1962a}) to define Quantum fields on curved
 spacetimes (e.g. \cite{Avis:1978tc,Keyl:1997cr,Radzikowski:1992eg,Verch:vs00,Sahlmann:2000ij,Brunetti:1999jn}).
 One may pass from these approaches to the $C^*$-algebraic in the same manner
 as in Minkowski spacetime.
 \\ For Wightman-fields
 in Minkowski spacetime it is known that the field algebra of a
 timelike tube is equal to the field algebra of the causal
 completion of this tube (\cite{Borchersr}). This may be derived
 (see \cite{Araki:1963}) from a mean value theorem by Asgeirsson (\cite{Asgeirsson:1936,Asgeirsson:1948}) and
 one of its consequences, namely that a solution to the wave
 equation vanishing in such a tube, vanishes in the causal
 completion of this tube (see \cite{CourantHilbert2}). As far as
 more general spacetimes are concerned the author showed recently
 that a similar result holds for stationary spacetimes and quite
 general free fields (\cite{Strohmaier:2000ye}). Namely the field
 algebra associated with a non-void  open set which is invariant
 under the time translation is equal to the quasilocal algebra.
 This result was used there to prove the Reeh-Schlieder property of
 vacuum-like states. The question arises whether such a result
 holds for more general spacetimes. As we will see this question is strongly
 related to the timelike Cauchy problem for second order hyperbolic
 partial differential equations. In the case of analytic coefficients on an analytic
 manifold Holmgrens' uniqueness theorem (see e.g.
 \cite{HormBook:1990}) states that there is unique continuation of
 solutions across any non-characteristic hypersurface. Relaxing
 however the condition of analyticity, counterexamples show that
 the Cauchy problem for timelike surfaces is ill posed in the
 general case of partial differential equations with smooth
 coefficients (see e.g. \cite{Hoerm:1975,Alinhac:1984} and
 references therein). Even for the wave operator
 $\square=\dzd{t}-\Delta$ in Minkowski spacetime it is known (see
 \cite{Alinhac:1995}) that there exists a smooth function $u$ such
 that there is a solution $\phi$ to the equation
 \begin{displaymath}
  \square \phi + u \phi = 0
 \end{displaymath}
 which has support equal to a half-space with timelike boundary.
 It was quite recently shown by Tataru in
 \cite{Tataru:1995} and further generalized by H\"ormander in
 \cite{Hoerm:1997} (see also \cite{Robbiano:1998})
 that under partial analyticity assumptions
 one still has unique continuation of solutions across
 non-characteristic surfaces.
 We give some global consequences of these unique continuation results.
 After defining an envelope of uniqueness $\mathcal{E}(\openset)$ of an
 open subset $\openset$ of the spacetime we show that a solution
 to a wave equation vanishing in $\openset$ vanishes in
 $\mathcal{E}(\openset)$. This envelope may be considerably larger than the original set and
 in particular in the case of small open neighbourhoods of inextendible
 timelike curves the envelope may coincide with the whole space.
 We investigate the consequences of these results for the real scalar
 field on curved spacetimes and show that the field algebra associated with the open set
 coincides with the field algebra associated with its envelope.
 The class of
 spacetimes considered here includes a large variety of physically
 relevant spacetimes, like cosmologically interesting
 Robertson-Walker spacetimes. We give an example of how the algebras
 associated with timelike curves can be explicitly given in curved
 spacetimes.

\section{Causal structure of spacetimes}

In the next two subsections we review the most important
definitions and results about the causal structure of spacetimes.
Details can be found in \cite{HawkingEllis}, \cite{Neill:1983} and
references therein. In the last two subsections we define what we
mean by the ``envelope'' of an open set. To do this we need to introduce
the space of timelike curve-segments endowed with an appropriate topology.

\subsection{Spacetimes and Causality}

We assume we are given a smooth connected manifold $M$ (Hausdorff
and second countable) of dimension $n \geq 2$ with a smooth metric
$g$ of Lorentzian signature. We assume in addition that $M$ is
oriented and time-oriented. In this case we say that $M$ is a spacetime.
For a subset $\mathcal{S}$ we
define the chronological future/past $I^{\pm}(\mathcal{S})$ of
$\mathcal{S}$ to be the set of points in $M$ that can be reached
by future/past directed timelike curves
\footnote{by curves we always mean maps from \underline {open} intervals in $\rz$ into
  $M$.}.
We define the future/past domain of dependence $D^{\pm}(\mathcal{S})$
of a subset $\mathcal{S} \subset M$
as the set of points $p$ in $M$ such that every past/future
inextendible causal curve through p intersects $\mathcal{S}$. The
domain of dependence $D(\mathcal{S})$ is the union
$D^+(\mathcal{S}) \cup D^-(\mathcal{S})$ (see \cite{GerochDD} for
a review). The causal complement $\openset^\perp$ of an open set
$\openset$ is the set of points which cannot be reached by causal
curves from $\openset$. The causal completion of $\openset$ is
$\openset^{\perp \perp}$.

\subsection{Globally hyperbolic manifolds}

A set $\mathcal{S}$ is called
achronal if every timelike curve intersects $\mathcal{S}$ at most once.
A spacelike hypersurface $\mathcal{C}$ is called Cauchy surface
if it is achronal and $D(\mathcal{C})=M$. In case there exists a
Cauchy surface the manifold $M$ is said to be globally hyperbolic.
An open subset $\openset \subset M$ is said to be globally hyperbolic
if $(\openset,g\vert_\openset)$ is globally hyperbolic as a spacetime in its
own right.
Note that for an achronal set $\mathcal{S}$ the open set
$\textrm{int}(D(\mathcal{S}))$ (if non-empty) is globally
hyperbolic.

\subsection{Space of timelike curve-segments}

We define curve-segments to be maps from compact intervals
in $\rz$ into $M$ which are restrictions of curves defined
on a larger interval. We say a curve-segment is timelike
if it is the restriction of a timelike curve.
For two points $p,q \in M$ let $C(p,q)$ the set of all smooth
timelike curve-segments $\gamma: [0,1] \to M$ with $\gamma(0)=p$
and $\gamma(1)=q$. These sets are commonly used to investigate properties
of globally hyperbolic spacetimes (see e.g. \cite{Choquet:1968,HawkingEllis}). 
Curve-segments in different parametrizations are identified.
We introduce on $C(p,q)$ a $C^1$-topology in the following
way: First we endow $M$ and $TM$ with complete Riemannian metrics. We
parameterize all curve-segments proportionally to their
arc-length in the metric on $M$. The
derivation of a curve-segment $\gamma$ with respect to this parameter
gives a curve-segment $\dot \gamma$ in $TM$. We now define a metric on
$C(p,q)$ in the following way
\begin{equation}
 d(\gamma_1,\gamma_2)=\sup_{t \in [0,1]} \textrm{dist}(\dot\gamma_1(t),\dot\gamma_2(t)),
\end{equation}
where $\textrm{dist}(x,y)$ denotes the Riemannian distance between
the points $x$ and $y$ in $TM$. One checks that the corresponding topology is
independent of the Riemannian metrics chosen on $M$ and $TM$.
\\ Given a smooth
timelike curve-segment $\gamma$ joining $p$ and $q$ we denote by
$C_0(p,q,\gamma)$ the connected component of $\gamma$ in $C(p,q)$.
We define the two sets
\begin{eqnarray}
 I(p,q):=\bigcup_{\tilde\gamma \in C(p,q)} \tilde \gamma((0,1)),\\
 I_0(p,q,\gamma):=\bigcup_{\tilde\gamma \in C_0(p,q,\gamma)} \tilde
 \gamma((0,1)),
\end{eqnarray}
i.e. the set of points which are met by the curves
$\tilde\gamma\vert_{(0,1)}$ with $\tilde \gamma$ in $C(p,q)$
or $C_0(p,q,\gamma)$ respectively.
One has the following properties
\begin{pro}\hfill
 \begin{enumerate}
  \item $I(p,q)$ coincides with $I^+(q) \cap I^-(p)$ in
        case $p$ is in the future of $q$.
  \item $I(p,q)$ and $I_0(p,q,\gamma)$ are open subsets of $M$.
  \item $I(p,q)$ and $I_0(p,q,\gamma)$ are invariant under conformal
        transformations of the metric.
 \end{enumerate}
\end{pro}\noindent
{\sl Sketch of Proof.}
 The definitions of $I(p,q)$ and $I_0(p,q,\gamma)$ depend only on the causal
 structure of the spacetime. Therefore, the sets are invariant under
 conformal transformations of the metric.
 The fact that $I_0(p,q,\gamma)$ and $I(p,q)$ are open follows from the
 causality properties of the tangent space and the properties of the
 exponential map (see in particular \cite{Neill:1983}, Lemma 33).
 If $p$ is in the future of $q$ and $x \in I^+(q) \cap I^-(p)$ then one can
 again use the exponential map $\textrm{exp}_x: T_x M \to \openset \subset M$
 to construct a smooth timelike curve through $x$ joining $p$ and $q$.
\hfill$\square$\\
In Minkowski space $I_0(p,q,\gamma)$ and $I(p,q)$ coincide,
this need not be the case however in general, even if one restricts the class
of spacetimes to the globally hyperbolic ones.
A simple expample is given
by the 2-dimensional cylinder $\rz \times S^1$ with metric $dt^2-d\phi^2$.
\begin{figure}
\centerline{
\includegraphics*[width=6cm]{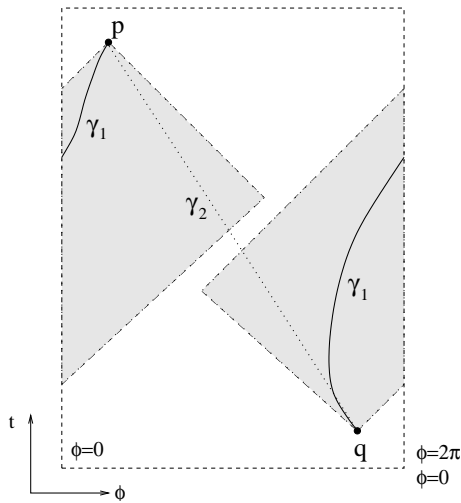}}
\vspace{0.1cm}
\caption{Right and left edge are identified, $I(p,q,\gamma_1) \not= I(p,q)$} \label{exc1}
\end{figure}
The curves $\gamma_1$ and $\gamma_2$ (see figure \ref{exc1}) cannot be
deformed continuously into one another and the set $I_0(p,q,\gamma_1)$ (the
shaded region) is different from $I(p,q)$ since it does not contain
$\gamma_2$.

For later considerations we need the
following lemma.
\begin{lem}\label{tube}
 Let $M$ be an n-dimensional spacetime and $p,q \in M$.
 Given a curve-segment $\gamma \in C(p,q)$ there is an open neighbourhood
 $\openset$ of $\gamma([0,1])$ such that the following holds
 \begin{itemize}
  \item There is a surjective local
      diffeomorphism\footnote{We say a map $f: M \to N$ is a local diffeomorphism if for each
      point $x \in M$ there is a neighbourhood $\openset \ni x$ such that
      $f\vert_\openset: \openset \to f(\openset)$ is a diffeomorphism.}
    $f : X \times B^{n-1} \to \openset$,
    where $B^{n-1}$ is the open unit-ball in $\rz^{n-1}$ and $X \subset \rz$
    a finite open interval.
  \item for each point $x \in B^{n-1}$ the curve $X \to M,\; t \to f(t,x)$
        is timelike.
 \end{itemize}
 If $\gamma_1 \in C(p,q)$ is sufficiently close to $\gamma$ in the $C^1$-topology we can choose
 $\openset$ and $f$ in such a way that the curve $X \to M,\; t \to f(t,0)$ is
 an extension of $\gamma_1$.
\end{lem}\noindent
 {\sl Sketch of Proof.}
 We extend the curve-segment $\gamma$ to a timelike curve
 $\tilde\gamma: I \to M$, where $I \subset \rz$
 is an open interval containing $[0,1]$. We choose another open interval
 $X$ in $I$ which is relatively compact and still contains $[0,1]$.
 Note that $\tilde \gamma$ is an immersion. This means that locally $\tilde \gamma$
 is an imbedding and we can form the conormal bundle $T_N^* I$ over $I$
 together with a natural immersion
 $\zeta$ from $T_N^* I$ into $T^*M$.
 We use an arbitrary Riemannian metric on $M$ to identify $TM$ and
 $T^*M$.
 Then the exponential map defines a smooth map $T^*M \supset \openset_0 \to M$,
 where $\openset_0$ is an open neighbourhood of the zero section.
 The map $j=\textrm{exp} \circ \zeta$ is well defined in a neighbourhood
 of the zero section in $T_N^* I \cong I \times \rz^{n-1}$. Moreover it has full rank at the zero
 section and the curve $\rz \supset t \to j(t,0)$ is timelike.
 Hence, there exists a neighbourhood $\openset_1 \subset T_N^* I$ of the zero
 section such that $j\vert_{\openset_1}$ is a local diffeomorphism and such
 that for all points $p \in \openset_1$ the push-forward $j_*(\partial_t)(p)$
 is timelike, if $t$ is a coordinate for $I$.
 Since $X \subset I$ is relatively compact in $I$
 there is an $\epsilon>0$ such that the bundle of open balls in
 $T_N^* X \subset T_N^* I$
 with radius $\epsilon$ is contained in $\openset_1$ and is relatively
 compact in $\openset_1$.
 Hence, $j$ restricts to a map $f$ from a set of
 the form $X \times B^{n-1}$
 onto some open neighbourhood of $\gamma([0,1])$.
 This map is a local diffeomorphism and moreover, by relative compactness of $X \times B^{n-1}$, 
 the Lorentzian length of the tangent vectors of the curves $\gamma_x: t \to f(t,x)$
 is uniformly bounded from below by some $\delta>0$.\\
 One shows that for a curve-segment $\alpha$ sufficiently close to $\tilde\gamma\vert_{X}$
 there is a unique curve-segment $\tilde \alpha: X \to \frac{1}{2}B^{n-1}$,
 such that $\alpha=f \circ \hat \alpha$, where $\hat \alpha$
 is the curve-segment
 $X \to X \times B^{n-1},\; t \to (t,\tilde\alpha)$.
 Given $\tilde\alpha$ we can define a map
 \begin{displaymath}
 h: X \times B^{n-1} \to X \times B^{n-1},\; (t,x) \to (t,\frac{1}{2}x+\tilde
 \alpha(t))
 \end{displaymath}
 and it is clear that $f_1:=f \circ h$ is an immersion.
 It is easy to check that the differential $df_1\vert_{p}$
 depends continuously on the choice of
 $\alpha$ and the point $p \in X \times B^{n-1}$.
 Remember that we had the length of the tangent vectors of the curves
 $\gamma_x$ uniformly bounded from below by a $\delta>0$.
 Therefore, for $\alpha$ sufficiently close to $\tilde\gamma\vert_X$,
 the curves $\gamma_{x,1}: t \to f_1(t,x)$ are timelike for all
 $x \in B^{n-1}$. We can choose $\alpha$ as an extension of $\gamma_1$.
 One checks that the open set $\openset:=\textrm{Im}(f_1)$ and the local
 diffeomorphism $f_1: X \times B^{n-1} \to \openset$ have all the desired
 properties.
 \hfill$\square$

\subsection{The envelope of an open set}
Suppose now we are given an open subset $\openset$ of $M$. We
define the envelope $\mathcal{E}(\openset)$ of $\openset$ to be
the smallest set containing $\openset$ with the following
properties:
\begin{enumerate}
 \item If an open subset $\mathcal{U} \subset M$ is globally hyperbolic
       and $\mathcal{S}$ is a Cauchy surface for $\mathcal{U}$
       which is contained in $\mathcal{E}(\openset)$ then
       $\mathcal{U} \subset \mathcal{E}(\openset)$.
 \item For each timelike curve-segment $\gamma \in C(p,q)$ such that $\gamma([0,1])$
       is contained in $\mathcal{E}(\openset)$ the set
       $I_0(p,q,\gamma)$ is contained in $\mathcal{E}(\openset)$.
\end{enumerate}
Our definition follows \cite{Thomas:1997ni} and \cite{Araki:1963}, where such an
envelope was defined for open sets in Minkowski spacetime,
instead of lines we use however arbitrary timelike curves.
Note that the envelope is well defined, since the intersection of two
sets with the listed properties has again these properties. One
checks that the causal completion $\openset^{\perp \perp}$ always
fulfills 1 and 2 and hence, $\mathcal{E}(\openset)
\subset\openset^{\perp \perp}$. Furthermore
$\mathcal{E}(\openset)$ is open, since if a set satisfies 1 and 2
this is true for the interior of this set as well.

\section{Unique continuation and the envelope of uniqueness}\label{ucp}

In this section $(M,g)$ will be an $n$-dimensional spacetime.
\begin{definition}
 We say a family of tensors $(f_i)$ is locally analytic in the time
 if for each point $p \in M$ there is an open neighbourhood $\openset$
 and a chart $u: \openset \to \mathcal{U} \subset \rz^n$ with coordinates
 $(x_0,x_1,\ldots,x_{n-1})$ mapping $p$ to $0$ such that
 \begin{itemize}
  \item in coordinates the local vectorfield $\partial_{x_0}$ is timelike.
  \item the components of the tensors $f_i$ in the coordinate basis
 viewed as functions on $\mathcal{U}$ are analytic in $x_0$ in a
 neighbourhood of 0.
 Hence, they are defined on
 $\{(z,x) \in \cz \times \rz^{n-1}; \vert z \vert <r, \vert x \vert <r \}$
 for some $r>0$.
 \end{itemize}
\end{definition}

If $H$ is a hypersurface defined as the zero set of a smooth local function
$h: M \supset \mathcal{U} \to \rz$ we say there is unique continuation
for a class of distributions $\mathcal{T}$ across
that hypersurface $H$
if the following is true for each point $x \in H$:
if there is a neighbourhood $\openset$ of $x$ in $\mathcal{U}$ such that
$g \in \mathcal{T}$ is zero in the open set $\{ y \in \openset ; h(y)<0 \}$
then $x$ is not in the support of $g$.
Specializing the results in \cite{Tataru:1995} and in particular in
\cite{Hoerm:1997} (the theorem and the remark in section 5) to the case of
second order differential equation with metric principal part, we get
that as soon as the coefficients of such a differential equation
are locally analytic in the time there is unique continuation of
distributional solutions across any non-characteristic hypersurface.
This result is local and we will study now the global consequences.

Denoting by $\nabla$ the Levi-Civita connection, the
wave operator $\square_g$ is given by $g^{ik} \nabla_i
\nabla_k\;$. The Klein-Gordon equation with mass $m \geq 0$ and
coupling $\kappa \in \rz$ reads
\begin{equation}
 (\square_g +m^2 +\kappa R) \psi = 0,
\end{equation}
where $R$ is the scalar curvature.
We treat a more general form of equation and show the following.

\begin{theorem}\label{vanishth}
 Let $\psi \in \mathcal{D}'(M)$ be a solution to the wave equation
 \begin{displaymath}
  (\square_g + a^k(x) \partial_k + V(x)) \psi = 0,
 \end{displaymath}
 where $a$ is a smooth vector field and $V$ a smooth potential.
 Assume that the family of tensors $(g, a, V)$ is locally analytic in
 the time.
 If $\psi$ vanishes in an open set $\openset$,
 then it vanishes in the envelope $\mathcal{E}(\openset)$ of this set.
\end{theorem}

\begin{proof}
 A distributional solution on a globally hyperbolic manifold $\mathcal{U}$
 vanishing in a neighbourhood of a Cauchy surface $\mathcal{S}$
 vanishes in $\mathcal{U}$. This is a general
 property of hyperbolic wave equations (see
 e.g. \cite{HawkingEllis, Friedlander, Leray}).
 We will show, that if a solution vanishes in a neighbourhood $\hat\openset$
 of $\gamma([0,1])$ for some $\gamma \in C(p,q),\;p,q \in M$ then it
 vanishes in the set $I_0(p,q,\gamma)$.
 Let $\mathcal{V}$ be the open subset of $C_0(p,q,\gamma)$ consisting
 of curves $\tilde\gamma$ such that $\textrm{supp}(\psi)$
 is disjoint from some open neighbourhood of $\tilde\gamma([0,1])$.
 We show that the boundary $\partial\mathcal{V}$ of $\mathcal{V}$ is empty.
 Thus $\mathcal{V}$ is open and closed and therefore coincides with $C_0(p,q,\gamma)$.
 Suppose the boundary $\partial\mathcal{V}$ of $\mathcal{V}$ were non-empty and
 let $\gamma_1$ be a curve-segment in $\partial\mathcal{V}$.
 This implies that there is a point $x \in \textrm{supp}(\psi)$
 which is met by $\gamma_1$ (see also figure \ref{proof}).
 \begin{figure}
  \centerline{\includegraphics*[width=6cm]{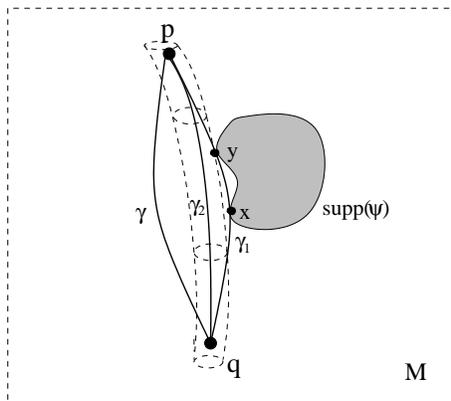}}
  \caption{Geometrical idea underlying the proof of theorem \ref{vanishth}}\label{proof}
 \end{figure}
 Since $\gamma_1 \in \partial\mathcal{V}$
 there is a curve-segment $\gamma_2 \in \mathcal{V}$ which is sufficiently
 close to $\gamma_1$, so that we can choose an open neighbourhood 
 $\openset_1$ of $\gamma_1([0,1])$ and a local diffeomorphism
 $f: \rz \supset X \times B^{n-1} \to \openset_1$ with the
 properties listed in lemma \ref{tube}, in particular such that
 $\gamma_2([0,1]) \subset f(X \times \{ 0 \} )$.
 We may even choose $f$ and $\openset_1$ such that $f$ has a continuous
 extension $\overline{f}$ to $\overline{X \times B^{n-1}}$
 with $\overline{f}(\partial X \times B^{n-1}) \subset \hat\openset$.
 Then there exists an open ball $B^{n-1}_r$ with radius
 $0 < r < 1$ such that
 $X \times B^{n-1}_r \cap f^{-1}(\textrm{supp}(\psi))=\emptyset$
 and $Y:=S_r \cap f^{-1}(\textrm{supp}(\psi))\not=\emptyset$ with
 $S_r:=X \times \partial B^{n-1}_r$.
 We take a point $y \in Y$ and a neighbourhood $\tilde \openset$
 of $y$ such that $f\vert_{\tilde \openset}$ is a diffeomorphism onto
 an open neighbourhood of $f(y)$. Because $f$ has the properties
 listed in lemma \ref{tube} (in particular the second property),
 the smooth hypersurface $f(S_r \cap \tilde \openset)$ is timelike and hence
 non-characteristic. Therefore there is unique continuation
 for solutions across this surface.
 The solution $\psi$ vanishes on $f((X \times B^{n-1}_r) \cap \tilde \openset)$
 and by unique continuation the point $f(y)$ cannot be in
 $\textrm{supp}(\psi)$ which is a contradiction.
 Hence, $\partial \mathcal{V}$ is empty.\\
 Since the envelope $\mathcal{E}(\openset)$ is the smallest set containing
 $\openset$ with the properties that have now been shown for the complement
 of $\textrm{supp}(\psi)$, it follows that 
 $\psi$ vanishes on $\mathcal{E}(\openset)$.
\end{proof}\noindent
In case the metric tensor is locally analytic in the time the
Klein-Gordon equation for mass $m$ and coupling $\kappa$ satisfies the
assumptions of the above theorem.

\begin{exa}
 If $(S,h)$ is a complete connected Riemannian manifold
 and  $f: I \to \rz^+$ is a smooth positive function on an interval $I \subset \rz$
 then the manifold $M:=I \times S$
 with metric $g:=dt \otimes dt - f(t) h$ is a globally hyperbolic spacetime.
 In case $S$ has constant curvature such spacetimes are called Robertson-Walker
 spacetimes. If $f$ is analytic the metric tensor is locally analytic in the
 time. Examples are the Friedmann models (see e.g. \cite{Neill:1983} for details).
\end{exa}

\section{The real scalar field on a globally hyperbolic spacetime}

 In this section we recall the construction of the real scalar field on a globally hyperbolic
 spacetime $M$.
 The Klein-Gordon operator
 for mass $m \geq 0$ and coupling $\kappa$ is:
 \begin{equation}
   P:=\square_g +m^2 + \kappa R.
 \end{equation}
 This operator acts on the real-valued smooth functions with compact support
 $C^\infty_0(M,\rz)$.
 It has unique advanced and retarded fundamental solutions (see \cite{Leray})
 $\Delta^\pm: C^\infty_{0}(M,\rz) \to C^\infty(M,\rz)$
 satisfying
 $$ P \Delta^\pm = \Delta^\pm P = \mathrm{id} \quad \textrm{on}
 \quad C^\infty_0(M,\rz)\;,$$
 $$ \textrm{supp}(\Delta^\pm f) \subset J^\pm(\textrm{supp}(f))\;.$$
 With $\Delta:=\Delta^+-\Delta^-$, $\hat\sigma(f_1,f_2):=\int_M f_1 \Delta(f_2) w$ defines
 an antisymmetric bilinear form on $C^\infty_{0}(M,\rz) \times C^\infty_{0}(M,\rz)$, where $w$
 is the pseudo-Riemannian volume form on $M$.
 Defining $\mathcal{W}:= C^\infty_{0}(M,\rz)/\textrm{ker}(\Delta)$ with quotient map
 $\eta$, the bilinear form $\sigma(\eta(f_1),\eta(f_2)):=\hat \sigma(f_1,f_2)$ on
 $\mathcal{W}$ is symplectic.
 The field algebra $\feld$ is defined to be the CCR-algebra
$\textrm{CCR}(\mathcal{W},\sigma)$ (see \cite{Kay:1993gr,Kay:1991mu,Bratteli2}).
This is the $C^*$-algebra generated by symbols $W(v)$ with
$v \in\mathcal{W}$ and the relations
\begin{eqnarray} \label{ccrrel}
  W(-v)=W(v)^*,\\
  W(v_1)W(v_2) = e^{-i \sigma(v_1,v_2)/2}W(v_1+v_2).
\end{eqnarray}
We define for each open subset $\openset \subset M$
the local field algebra $\feld(\openset) \subset \feld$ to be the closed $*$-subalgebra
generated by the symbols $W(\eta(f))$ with $f \in C^\infty_0(\openset,\rz)$.\\
It was shown in \cite{Dimock:1980hf} that there is a canonical representation $\tau$
of the group of isometries $G$ of $M$ by Bogoliubov automorphisms of $\feld$
and the net
$\openset \to \feld(\openset)$ has the following properties:
\begin{enumerate}
  \item Isotony: $\openset_1 \subset \openset_2$ implies
    $\mathcal{F}(\openset_1) \subset \mathcal{F}(\openset_2)$.
  \item Causality: if $\openset_1 \subset \openset_2^\perp$, then
        $[\feld(\openset_1),\feld(\openset_2)]=\lbrace 0 \rbrace$.
  \item Covariance: $\tau(q) \mathcal{F}(\openset)=\mathcal{F}(q \openset) \quad \forall q \in G$.
\end{enumerate}
Moreover, $\feld$ is the quasilocal algebra of the net
$\openset \to\mathcal{F}(\openset)$.\\
Assume that we are given a scalar product $\mu$ on $\mathcal{W}$
which dominates $\sigma$, i.e. satisfies the estimate
\begin{equation} \label{dom}
 \vert \sigma(v_1,v_2) \vert^2 \leq 4 \mu(v_1,v_1) \mu(v_2,v_2) \quad
 v_1,v_2 \in \mathcal{W}.
\end{equation}
In this case the linear functional $\omega_\mu : \feld \to \cz$, defined by
\begin{equation}
  \omega_\mu(W(v)):=e^{-\mu(v,v)/2} \quad v \in \mathcal{W},
\end{equation}
is a state (see \cite{Kay:1993gr,Kay:1991mu,Bratteli2}). The states over $\feld$ which can be realized in this
way are called quasifree states.
A quasifree state $\omega_\mu$ gives rise to a one particle structure (Proposition 3.1 in \cite{Kay:1991mu}), that
is a map $K_\mu: \mathcal{W} \to H_\mu$ to some complex Hilbert space $H_\mu$, such
that
\begin{enumerate}
  \item the complexified range of $K_\mu$, (i.e. $K_\mu \mathcal{W} + i K_\mu \mathcal{W}$), is
        dense in $H_\mu$,

  \item $\langle K_\mu v_1,K_\mu v_2 \rangle=\mu(v_1,v_2)+\frac{i}{2}\sigma(v_1,v_2)$.
\end{enumerate}
This structure is unique up to equivalence.
A one particle structure $(K_\mu,H_\mu)$ for a quasifree state allows one to
construct the GNS-triple
$(\pi_{\omega_\mu},\Hilbert_{\omega_\mu},\Omega_{\omega_\mu})$ explicitly
(see \cite{Kay:1991mu,Kay:1993gr,Bratteli2}).
Namely, one takes $\Hilbert_{\omega_\mu}$ to be the bosonic Fock
space over $H_\mu$ with Fock vacuum $\Omega_{\omega_\mu}$, and defines the
representation by
$\pi_{\omega_\mu}(W(v))=\textrm{exp}(-(\overline{\hat a^*(K_\mu v)-\hat a(K_\mu v)}))$,
where $\hat a^*(\cdot)$ and $\hat a(\cdot)$ are the usual creation and
annihilation operators.
One has the following (see e.g. \cite{Araki:1982cc}, Proposition 3.4~(iii)):
\begin{pro} \label{Bosedens}
 Let $\omega_\mu$ be a quasifree state over the $C^*$-algebra
 $\feld = \textrm{CCR}(\mathcal{W},\sigma)$ and let
 $(\pi_{\omega_\mu},\Hilbert_{\omega_\mu},\Omega_{\omega_\mu})$ be its
 GNS-triple.
 If $V \subset \mathcal{W}$ is a subspace which is dense in $\mathcal{W}$
 in the topology defined by $\mu$, then the $*$-algebra generated by the set
 \begin{displaymath}
   \lbrace \pi_{\omega_\mu}(W(v)), v \in V \rbrace \subset \pi_{\omega_\mu}(\feld)
 \end{displaymath}
 is strongly dense in the von Neumann algebra $\pi_{\omega_\mu}(\feld)''$.
\end{pro}

\begin{definition}
 Let $\feld$ be the field algebra of the real scalar field
 with mass $m \geq 0$ and coupling $\kappa$.
 We call a quasifree state $\omega_\mu$ over $\feld$ continuous
 if the 2-point function $w_2(\cdot,\cdot):=\langle K_\mu \eta(\cdot), K_\mu \eta(\cdot) \rangle$
 is a distribution in $\mathcal{D}'(M \times M)$.
\end{definition}\noindent
Given a continuous quasifree state $\omega_\mu$ we can construct the net
of von Neumann algebras
$\openset \to \hat\feld(\openset):=\pi_{\omega_\mu}(\feld(\openset))''$.
This assignment is isotone, causal and covariant, and there is a unique
$\sigma$-weakly-continuous representation $\hat \tau$ of $G$ on
$\hat\feld=\pi_{\omega_\mu}(\feld)''$ which extends $\tau$.
The net gives rise to a quantum
field theory on $M$.

\section{The local structure of the real scalar field}\label{cucp}

\begin{theorem}\label{mainth}
 Let $M$ be a globally hyperbolic spacetime
 with a metric which is locally analytic in the time.
 Let $\openset \to \feld(\openset)$ be the net of $C^*$-algebras
 for the real scalar field on $M$ with mass $m \geq 0$ and coupling $\kappa$.
 If $\omega$ is a continuous quasifree
 state over the quasilocal algebra $\feld$, then
 for each open set $\openset$ the von Neumann algebra
 $\hat\feld(\openset)=\pi_{\omega} (\feld (\openset))''$ is equal to the von
 Neumann algebra $\hat\feld(\mathcal{E}(\openset))=\pi_{\omega} (\feld (\mathcal{E}(\openset)))''$, i.e.
 the local field algebras of an open set and its envelope coincide.
\end{theorem}

\begin{proof}
 Let $\mu$ be the scalar product on $\mathcal{W}$ inducing the state $\omega$.
 By proposition \ref{Bosedens} it is sufficient to show that $\eta(C_0^\infty(\openset,\rz))$
 is $\mu$-dense in $\eta(C_0^\infty(\mathcal{E}(\openset),\rz))$.
 We show that a $\mu$-continuous linear form $\hat\psi$ on $\mathcal{W}$ vanishing on
 $\eta(C_0^\infty(\openset,\rz))$ vanishes on the set
 $\eta(C_0^\infty(\mathcal{E}(\openset),\rz))$.
 Note that $\psi:=\hat\psi(\eta(\cdot))$ is a real-valued
 distribution in $\mathcal{D}'$
 and a solution to the Klein-Gordon equation. By assumption $\psi$
 vanishes in $\openset$ and by theorem \ref{vanishth}
 it vanishes in $\mathcal{E}(\openset)$.
 Hence, the theorem is proved.
\end{proof}

\begin{rem}
 Of course the same conclusion holds on general spacetimes
 with metric locally analytic in the time
 as soon as the field operator satisfies the Klein-Gordon equation.
 Here we specialized to the case of a globally hyperbolic spacetime
 since we gave the construction of the field only in this case.
\end{rem}
\noindent
Given a local net of von Neumann algebras $\openset \to \hat\feld(\openset)$
on a spacetime $M$ one may define the local algebras associated to
curves (see \cite{Wollenberg:1998}).
For a  curve $\gamma: I \to M$ which is contained in a compact subset
$\mathcal{K} \subset M$ we define
the algebra of the curve to be the von Neumann algebra
\begin{equation}
 \hat\feld(\gamma):= \bigcap_{\openset \supset \gamma(I)} \hat\feld(\openset).
\end{equation}
An immediate consequence of our theorem is
\begin{cor}
  Let the assumptions of theorem \ref{mainth} be fulfilled.
  Let $\openset \to \hat\feld(\openset)$ be the corresponding net
  of von Neumann algebras (see the end of the previous section).
  If $\gamma: (0,1) \to M$ is a timelike curve which can be continued
  to a curve segment $\tilde \gamma: [0,1] \to M$ with endpoints $p$
  and $q$. Then the algebra $\hat\feld(\gamma)$ coincides with
  $\hat \feld(I_0(p,q,\tilde \gamma))$. 
\end{cor}
\noindent
This shows that the algebra of a timelike curve coincides with the algebra
of a neighbourhood of this curve. This was conjectured in
\cite{Wollenberg:1998}, p.239 to hold for globally hyperbolic spacetimes and
our result may be seen as a partial positive answer.
Another application is
\begin{exa}
 If $(S,h)$ is a complete connected Riemannian manifold the manifold
 $M:=\rz \times S$ with metric $g:=dt \otimes dt -h$ is a globally hyperbolic
 Lorentzian manifold. Such manifolds are called ultrastatic.
 If $\openset \subset S$ is non-void, then one can show
 that the envelope of the set $\rz \times \openset$ coincides with $M$.
 The same holds for any metric that is conformally equivalent
 to $g$. If $f$ is a positive function on $M$ which is locally analytic
 in $t$ and $(M,\tilde g)$ is the Lorentzian manifold with metric
 $\tilde g:=f \cdot g$ then the local algebra
 $\hat \feld (\rz \times \openset)$ of the real scalar field on $(M, \tilde g)$
 coincides with the quasilocal algebra
 $\hat\feld$ whenever $\openset$ is non-void.
\end{exa}

\section{Acknowledgements}
The author would like to thank Prof.~M. Wollenberg, Dr.~R. Verch and
Prof.~H.J. Borchers for useful discussions and comments.
This work was supported by the
Deutsche Forschungsgemeinschaft within the scope of the postgraduate
scholarship programme ``Graduiertenkolleg Quantenfeldtheorie'' at the
University of Leipzig.

\end{document}